\documentclass[
aps,%
12pt,%
oneside,%
nobibnotes,%
nofootinbib,% In the current version of REVTeX, when this option is on,
superscriptaddress,%
amsmath,amssymb,
noshowpacs,%
tightenlines,
twocolumn]%
{revtex4}

\usepackage{graphics}
\usepackage{graphicx}
\usepackage{epsf} 
\usepackage{amsmath}

\def\del{\partial}

\def\beas{\begin{eqnarray}}
\def\eeas{\end{eqnarray}}
\def\bea{\beas}
\def\eea{\eeas}
\def\be{\begin{equation}}
\def\ee{\end{equation}}
\def\refitem#1{}
\def\setcaptionmargin#1{}
\def\onelinecaptionsfalse{}

\begin{document}
% \selectlanguage{russian} % For a paper in Russian
%\selectlanguage{english} % For a paper in English
\hyphenation{fourier-trans-formed sin-gle di-a-gram anoma-lous mag-netic mo-ment}

%\preprint{APS/123-QED}

%\begin{widetext}
\phantom{0}
\vspace{-0.4in}
\hspace{4in}\parbox{1.5in}{ \leftline{JLAB-THY-04-219}
                \leftline{WM-04-104}
%			             \leftline{nucl-th/04?????}
\vspace{1.1in}}
%\end{widetext}

\title{Nonperturbative dynamics of scalar field theories through the Feynman-Schwinger representation}
%\title{Feynman-Schwinger representation: numerical results in scalar field theories}
\date{\today}

\author{Cetin Savkli}
\affiliation{Lockheed Martin Space Operations, 7500 Greenway Center Drive, Greenbelt, MD 20770, USA}
\author{Franz Gross}
\affiliation{Jefferson Lab, 12000 Jefferson Ave,  Newport News, VA 23606, USA}
\author{John Tjon}
\affiliation{Jefferson Lab, 12000 Jefferson Ave,  Newport News, VA 23606, USA}
\affiliation{Department of Physics, University of Maryland, College Park, MD 20742, USA}

\begin{abstract}
In this paper we present a summary of results obtained for scalar field theories  using the Feynman-Schwinger (FSR) approach. Specifically, scalar QED
and $\chi^2\phi$ theories are considered. The motivation behind the applications 
discussed in this paper is to use the FSR method as a rigorous tool for testing 
the quality of commonly used approximations in field theory. Exact calculations in
a quenched theory are  presented for one-, two-, and three-body bound states.  Results obtained 
indicate that some of the commonly used approximations, such as Bethe-Salpeter ladder summation 
for bound states and the rainbow summation for one body problems, produce 
significantly different results from those obtained from the FSR approach.  We find that more accurate results can be obtained using other, simpler, approximation schemes.  

\end{abstract}

\pacs{Valid PACS appear here}% PACS, the Physics and Astronomy
\maketitle

\section{Introduction}
In the study of hadronic physics one has to face the problem of determining 
the quantum dynamical properties of physical systems in which the interaction 
between the constituents is of a nonperturbative nature. In particular,
such systems support bound states and clearly nonperturbative
methods are needed to describe their properties.
Assuming that such systems can be described by a field theory,
one has to rely on some approximation scheme. One common 
approximation is known as perturbation theory. Perturbation theory involves 
making an expansion in the coupling strength of the interaction. 
The Green's function in field theory can be expanded in powers of 
the coupling strength. 
In order to be able to obtain a bound state result one must sum the 
interactions to all orders. Most practical calculations to date have
been done within the Bethe-Salpeter framework, where the resulting 
kernel is perturbatively truncated. In this paper we will
discuss another method, which is based on the path integral
formulation of Feynman  and Schwinger~\cite{Feynman:1950ir,Schwinger:1951nm}.
It was initiated by Simonov and collaborators~\cite{Simonov:1988rn,Simonov:1989mj,Simonov:1991re,Simonov:1993kp,Simonov:2002vv,Simonov:2002we,Simonov:2002zj}
in their study of quantum chromo-dynamics (QCD).

With the discovery of QCD, nonperturbative calculations in field 
theory have become even more essential. It is known that 
the building blocks of matter, quarks and gluons, only exist in bound states. 
Therefore any reaction that involves quarks will necessarily involve bound 
states in the initial and/or final states. This implies that even at high 
momentum transfers, where QCD is perturbative, formation of quarks into a 
bound state necessitates a nonperturbative treatment. Therefore it is 
essential to develop new methods for doing nonperturbative calculations in 
field theory.

The plan of this article is as follows. In the following section a 
brief review of the particle trajectory method in field theory will 
be given. 
The Feynman-Schwinger representation will be introduced through applications to
scalar fields. In particular, the emphasis will be on comparing various 
nonperturbative results obtained by different methods. It will be 
shown with examples 
that nonperturbative calculations are interesting and exact nonperturbative 
results could significantly differ from those obtained by approximate 
nonperturbative methods. 

Results for scalar quantum electrodynamics are discussed in 
section 3. In particular, a cancellation is found between the vertex corrections and 
self-energy contributions, so that the exact result can be described
by essentially the sum of only the generalized exchange ladders. In 
section 4 the bound states are obtained for a $\chi^2 \phi$
theory in a quenched approximation. Exact results are presented for the binding energies of
two- and three-body bound states and compared with relativistic quasi-potential 
predictions. Section 5 deals with the stability of the $\chi^2
\phi$ theory. It is argued that the quenched theory does not suffer from
the well known instability of a $\phi^3$ theory. The paper closes with
some concluding remarks.
 
\section{Feynman-Schwinger representation}

Nonperturbative calculations can be divided into two general categories: (i) integral equations,  and (ii) path integrals.
Integral equations have been used for a long time to sum interactions 
to all orders with various approximations~\cite{Salpeter:1951sz,Gross:1991te,Tiemeijer:1994bj,Savkli:1999me}.
In general a complete solution
of field theory to all orders can be provided by an infinite set of 
integral equations relating vertices and propagators of the theory to each 
other. However solving an infinite set of equations is beyond our reach 
and usually integral equations are truncated by various assumptions about
the interaction kernels and vertices. 
The most commonly used integral equations are those that deal
with few-body problems. The Bethe-Salpeter equations~\cite{Salpeter:1951sz} are the
starting point of those investigations.
Approximations schemes have extensively been studied,
where in addition to solutions of the Bethe-Salpeter equation~\cite{Wright:1967bs,Nakanishi:1969ph,Nakanishi:1988hp,Nieuwenhuis:1996bs,Savkli:1998kz}
also 3-dimensional reductions~\cite{Salpeter:1952qp,Logunov:1963yc,Blankenbecler:1966gx,Gross:1969rv,Gross:1982nz,Wallace:1989nm}
 have been explored for the N-particle free particle Greens' function. 
In most calculations the ladder approximation has been used for
the kernel of the resulting equations. Issues of convergence of
these schemes remain. Therefore it is important to have exact
solutions of field theory models available to test these
approximations. One promising way to reconstruct exact solutions
of field theory is the path integral method.  

Path integrals provide a systematic method for summing interactions to all orders. 
The Green's function in field theory is given by the path integral expression: 
\begin{widetext}
\beas
\langle 0| T[\phi(x_1)\phi(x_2)\cdots \phi(x_n)]|0\rangle&=&\frac{\biggl\lmoustache[{\cal D}\phi]\,\phi(x_1)\,\phi(t_2)\cdots \phi(x_n)\,{\rm exp}\biggl[i\int d^4x\,L(x)\biggr]}{\biggl\lmoustache[{\cal D}\phi]\,{\rm exp}\biggl[i\int d^4x\,L(x)\biggr]}. \label{eq1}
\eeas
\end{widetext}
While path integrals provide a compact expression for the exact 
nonperturbative result for propagators, evaluation of the path integral 
is a nontrivial task. 

In general, field theoretical path integrals must be evaluated by 
numerical integration methods, such as Monte-Carlo integration.
The best known numerical integration method is lattice gauge 
theory. Lattice gauge theory involves a discretization of 
space-time and numerical integrations over field configurations are 
carried out using Monte-Carlo techniques. 
A more efficient method of performing path 
integrals in field theory has been proposed and it consists of
explicitly integrating out the fields. It has been demonstrated to be
highly successful  for the case of simple scalar 
interactions~\cite{Simonov:1993kp,Nieuwenhuis:1994mc,Nieuwenhuis:1995sq,Nieuwenhuis:1996mc,Savkli:1999rw,Savkli:1999ui,Savkli:1999gq,Gross:2001ha,Savkli:2002fj,Savkli:2000hw}. 
This method is known as Feynman-Schwinger representation (FSR).
Through applications of the FSR, the importance of {\em exact} 
nonperturbative calculations will be shown with explicit examples.

The basic idea behind the FSR approach is to transform the field theoretical path integral (\ref{eq1}) into a quantum mechanical path integral over {\it particle trajectories\/}.   When written in terms of trajectories, the exact results decompose into separate parts, and permit us to study the individual role and numerical size of exchange, self-energy, and vertex corrections.   This, in turn, alows us to study different approximations to field theory, and, in some cases. prove new results.

To illustrate these ideas, we consider the
application of the FSR technique to scalar QED. The Minkowski metric 
expression for the scalar QED Lagrangian in Stueckelberg form is given by
\beas
{\cal L}_{SQED}&=&-m^2\chi^2-\frac{1}{4}F^2+\frac{1}{2}\mu^2A^2\nonumber\\
&&+(\partial_\mu-ieA_\mu)\chi^*(\partial^\mu+ieA^\mu)\chi
\nonumber\\
&&-
\lambda\frac{1}{2}(\partial\cdot A)^2\, ,
\eeas
where $A^\mu$ is the gauge field of mass $\mu$, $\chi$ is 
the charged field of mass $m$, $F^{\mu\nu}=\del^\mu A^\nu-\del^\nu A^\mu$ is the gauge field tensor, and, for example, $A^2=A_\mu A^\mu$. The presence of a mass term for the exchange 
field breaks gauge invariance, and was introduced in order 
to avoid infrared singularities that arise when the theory is applied in 0+1 dimensions. 
For dimensions larger than $n=2$ the infrared 
singularity does not exist and therefore the limit $\mu\rightarrow~0$ can be 
safely taken to insure gauge invariance.    

The path integral is to be performed in Euclidean metric. Therefore we perform
a Wick rotation:
\beas
{\rm exp}\biggl[i\int d^4x\,{\cal L}_M\biggr]\longrightarrow {\rm exp}\biggl[-\int d^4x\,{\cal L}_E\biggr] .
\eeas
The Wick rotation for coordinates is obtained by
\bea
&&x_0\rightarrow -ix_0,\nonumber\\
&&\partial_0=\frac{\partial}{\partial x^0}\rightarrow i\partial_0
\, .
\eea
The transformation of field $A$ under Wick rotation is found by noting 
that under a gauge transformation:
\bea
A_\mu&\rightarrow&A_\mu+\partial_\mu\Lambda\, .
\eea
Then, under a Wick rotation,
\bea
A_0&\rightarrow&iA_0\, ,
\eea
and the Wick rotated Lagrangian for SQED becomes
\bea
{\cal L}_{SQED}&=&\chi^*\biggl[m^2-\partial^2-2ieA\cdot \partial-ie\partial\cdot A\nonumber\\
&&\qquad+e^2A^2\biggr]\chi+{\cal L}_{A}\, .
\eea
The exchange field part of the Lagrangian is given by
\bea
{\cal L}_{A}&\equiv&\frac{1}{2}A_\mu(\mu^2g_{\mu\nu}-\lambda\partial_\mu\partial_\nu)A_\nu+\frac{1}{4}F^2
\nonumber\\
&=&\frac{1}{2}A_\mu\Bigl[(\mu^2-\Box )g_{\mu\nu}
\nonumber\\
&&\qquad+(1-\lambda)\partial_\mu\partial_\nu\Bigr]A_\nu \, .
\eea
We employ the Feynman gauge $\lambda=1$ which yields
\bea
{\cal L}_{A}&=&\frac{1}{2}A_\nu(\mu^2-\Box )A_\nu \, .
\eea
The two-body Green's function for the transition from an initial 
state~$\Phi_{i}(x,\bar x)$ to final state $\Phi_{f}(y,\bar y)$ is given by
\bea
G(y,\bar{y}|x,\bar{x})=N\int {\cal D}\chi^*\int {\cal D}\chi\int {\cal D}A\,
\nonumber\\
\times\Phi^*_f\Phi_i\,e^{-S_E},
\label{g0.sqed}
\eea
where 
\be
S_E=\int d^4x \,\,{\cal L}_{SQED},
\ee
and a gauge invariant 2-body state $\Phi$ is defined by
\bea
\Phi(x,\bar{x})&=&\chi^*(x)U(x,\bar{x})\chi(\bar{x}) .
\eea 
The gauge link $U(x,y)$ which insures gauge invariance of bilinear product of 
fields is defined by  
\begin{equation}
U(x,y)\equiv {\rm exp} \left[-ie\int_x^ydz\,A(z)\right].
\end{equation} 
One can easily see that under a local gauge transformation
\bea
\chi(x) & \rightarrow &{\rm e}^{ie\Lambda(x)}\chi(x)
\nonumber\\
A_\mu(x) & \rightarrow & A_\mu(x)+\partial_\mu\Lambda(x),
\eea
\begin{figure}
\setcaptionmargin{5mm}
%\onelinecaptionsfalse
\includegraphics[width=2.4in]{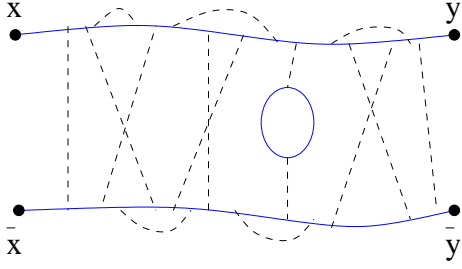}
%\includegraphics[width=2.4in]{unquenched}
%captionstyle{normal}
\caption{The dashed lines represent exchanges of the gauge field with mass $\mu$ and the solid lines the propagation of the matter fields with mass $m$.  Note the matter loop in one of the middle exchanges.  All loops of this kind are neglected in the quenched approximation (when $\det S=1$).}
\label{unquenched.fig} 
\end{figure}
$\Phi_{i}(x,\bar{x})$ remains gauge invariant
\begin{widetext}
\bea
\Phi(x,\bar{x})&\rightarrow&{\rm exp}\biggl[\underbrace{-ie\Lambda(x)+ie\Lambda(\bar{x})-ie\biggl\lmoustache_x^{\bar{x}}dz_\mu\,\partial_\mu\Lambda}_0\biggr]\chi^*(x)U(x,\bar{x})\chi(\bar{x})
=\Phi(x,\bar{x}).
\eea
Performing path integrals over $\chi$ and $\chi^*$ fields in Eq.~(\ref{g0.sqed})
one finds
\bea
G(y,\bar{y}|x,\bar{x})&=&N\bigg\lmoustache {\cal D}A\, ({\rm det}S)\,\,U(x,\bar{x})U^*(y,\bar{y})S(x,y)S(\bar{x},\bar{y})\,e^{-S[A]}\, ,
\label{green0}
\eea
\end{widetext}
where the interacting 1-body propagator $S(x,y)$ is defined by
\bea
S(x,y)&\equiv& \langle y\,|\,\frac{1}{m^2+H(\hat{z},\hat{p})}\,|\,x\rangle\label{s0}
\eea
with
\bea
H(\hat{z},\hat{p})&\equiv& (\hat{p}+ieA(\hat{z}))^2\label{hami}.
\eea
The Green's function Eq.~(\ref{green0}) includes contributions 
coming from all possible interactions. The determinant in Eq.~(\ref{green0}) 
accounts for all matter ($\chi\bar\chi$) loops.  Setting this determinant equal to unity (${\rm det} S\rightarrow 1$, referred to as the quenched approximation) eliminates all contributions from these loops (illustrated in Fig.~\ref{unquenched.fig})  and greatly simplifies the calculation.

Analytical calculation of the path integral over the gauge field $A$ in Eq.~(\ref{green0}) seems difficult due to the nontrivial $A$ dependence in $S(x,y)$. In more complicated theories, such as QCD, integration of the gauge field,
as far as we know, cannot be done analytically. Therefore, in QCD, the only 
option is to do the gauge field path integral by using a brute force method.  Here one usually introduces a  discrete space-time lattice, and integrates over the values of the field components at each lattice site . However, for the simple scalar QED 
interaction under consideration, it is in fact possible to go further
and eliminate the path integral over the field $A$. In order to be able to carry out
the remaining path integral over the exchange field $A$ it is desirable to 
represent the interacting propagator in the form of an exponential. This can 
be achieved by using a Feynman representation for the interacting propagator. 
The first step involves the exponentiation of the denominator in Eq.~(\ref{s0}):
\bea
S(x,y)=\int_0^{\infty}\!\!\!ds\,e^{-s m^2}\langle y|{\rm exp}[-sH] |x\rangle \, .
\label{sqeds}
\eea
This expression is similar to a quantum mechanical propagator with $s=it$ 
and $H$ a Hamiltonian which is a covariant function of 4-vector momenta and coordinates.
It is known how to represent a quantum 
mechanical propagator as a path integral.  The representation is in terms of the Lagrangian, and a covariant Lagrangian can easily be obtained from the Hamiltonian (\ref{hami})
\begin{figure}
\setcaptionmargin{5mm}
%\onelinecaptionsfalse
\includegraphics[width=2.0in]{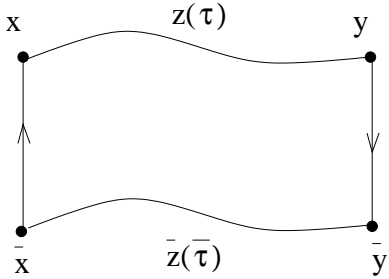}
%\includegraphics[width=2.0in]{wilsonloop}
%captionstyle{normal}
\caption{ The contour $C$, known as a Wilson loop, that arises in Eq.~(\ref{eqW}). }
\label{wilsonloop.fig} 
\end{figure}
\begin{widetext}
\bea
&&H(\hat{z},\hat{p})=(\hat{p}+ieA(\hat{z}))^2\,\,\Longrightarrow\,\,L(z,\dot{z})=\frac{\dot{z}^2}{4}-ie\,\dot{z}\cdot A(z)\, .
\eea
Using this Lagrangian, the path integral representation for the interacting propagator becomes
\bea
S(x,y)&=&\bigg\lmoustache_0^\infty ds\,\bigg\lmoustache\,({\cal D}z)_{xy}\,\,{\rm exp}\biggl[-sm^2-\frac{1}{4}\int_0^s d\tau_s\, \dot{z}^2(\tau_s)-ie\int_0^s d\tau_s\, \dot{z}A(z(\tau_s))\,\,\biggr],
\eea
where $z_i(\tau_s)$ is a particle trajectory which is a parametric function of the parameter $\tau_s$, with $s\ge\tau_s\ge0$ and endpoints $z_i(0)=x_i$, $z_i(s)=y_i$, and $i=1$ to 4. This 
representation allows one to perform the remaining path integral over the 
exchange field $A$. The final result for the two-body propagator 
involves a quantum mechanical path integral that sums up contributions coming 
from all possible {\em trajectories} of the two charged {\em particles}
\bea
G&=&-\bigg\lmoustache_0^\infty 
ds\,\, 
\bigg\lmoustache_0^\infty 
d\bar{s}\,\, \bigg\lmoustache\, 
({\cal D}z)_{xy}\,
\bigg\lmoustache\, 
({\cal D}\bar{z})_{\bar{x}\bar{y}}\,\, 
e^{ -K[z,s]-K[\bar{z},\bar{s}] }    
\langle W(C)\rangle,
\label{gfin}
\eea
where the parameter $\tau_s$ is rescaled, so that $\tau_s=s\tau$, the  kinetic term $K$ is defined by
\bea
K[z,s]&=&m^2s+\frac{1}{4s}\int_0^1 d\tau \,\dot{z}^2(\tau)\, ,
\eea
and the Wilson loop average $\langle W(C)\rangle$ is given by
\bea
\langle W(C)\rangle&\equiv&\int {\cal D}A\,{\rm exp}\biggl[-ie\oint_Cdz\,A(z)-\frac{1}{2}\int d^4z\,A(z)(\mu^2-\del^2)A(z)\biggr], \label{eqW}
\eea
where the contour of integration $C$ (shown in Fig.~\ref{wilsonloop.fig}) follows a clockwise trajectory $x\rightarrow y\rightarrow\bar{y}\rightarrow \bar{x}\rightarrow x$ as parameters $\tau$, and 
$\bar{\tau}$ are varied from 0 to $1$. 
The $A$ integration in the definition of the Wilson loop average is 
of standard gaussian form and can be easily performed to obtain
\bea
\langle W(C)\rangle&=&{\rm exp}\,\biggl[-\frac{e^2}{2}\int_C\,dz_\mu\,\int_C\,d\bar{z}_\nu \,\Delta_{\mu\nu}(z-\bar{z},\mu)\biggl],\label{v.sqed}\\
\Delta_{\mu\nu}(x,\mu)&=& g_{\mu\nu}\bigg\lmoustache \frac{d^4p}{(2\pi)^4}\frac{e^{ip x}}{p^2+\mu^2}\, .
\label{delta1}
\eea
When it is necessary to regulate the ultraviolet singularities in (\ref{delta1}),  a double Pauli-Villars subtraction will be used, so that (\ref{delta1}) will be replaced by 
\bea
\Delta_{\mu\nu}(x,\mu)&=& g_{\mu\nu}\bigg\lmoustache \frac{d^4p}{(2\pi)^4}
\frac{e^{ip x}
(\Lambda_1^2-\mu^2)(\Lambda_2^2-\mu^2)}{(p^2+\mu^2)(p^2+\Lambda_1^2)(p^2+
\Lambda_2^2)}\, . \label{delta2}
\eea
\end{widetext}
Through the results given in Eqs.~(\ref{gfin}), (\ref{v.sqed}), and either (\ref{delta1}) or (\ref{delta2}), the path integration 
expression involving {\em fields} has been transformed into a path integral 
representation involving {\em trajectories} of particles.

Equation~(\ref{gfin}) has a very nice physical interpretation.  The term $\Delta_{\mu\nu}(z_a-z_b,\mu)$ describes the propagation of gauge field
interations between any two points on the particle trajectories, and the
appearance of these interaction terms in the exponent means that the
interactions are summed to all orders with arbitrary ordering of the points
on the trajectories.  Self-interactions come from terms with the two points
$z_a$ and $z_b$ on the {\it same\/} trajectory, generalized ladder
exchanges arise if the two points are on {\it different\/} trajectories, and
vertex corrections arise from a combination of the two.   
Because the  particles forming the two-body bound state carry opposite charges, it follows that the self energy and exchange contributions have different signs. 

The bound state spectrum can be determined from the spectral decomposition of the two-body Green's function
\be
G(T)=\sum_{n=0}^{\infty}c_ne^{-m_nT},
\ee
where $T$ is defined as the average time between the initial and final states
\be
T\equiv\frac{1}{2}(y_4+\bar{y}_4-x_4-\bar{x}_4).
\label{tdef}
\ee
In the limit of large $T$, the ground state mass is given by
\begin{equation}
m_0=-\lim_{T\rightarrow\infty}\frac{d}{dT} {\rm ln}[G(T)]=-\frac{\int {\cal D}Z S'[Z]e^{-S[Z]}}{\int {\cal D}Z e^{-S[Z]}},
\label{groundstate}
\end{equation}

\section{Scalar quantum electrodynamics}
%Comparison of the exact FSR results with approximate nonperturbative methods}

In this section we take a closer look at 1-body mass pole calculations
for the case of SQED.
Two popular methods frequently used to find the dressed mass of a particle are to do a simple bubble summation,  or to solve the 1-body Dyson-Schwinger
equation in rainbow approximation. It is interesting to compare results 
given by the bubble summation and the Dyson-Schwinger with the exact FSR 
result.  
Below we first give a quick overview of how dressed masses can be obtained in 
bubble summation and the Dyson-Schwinger equation approaches.  For technical simplicity, the 1-body discussion will be limited to 0+1 dimension.

The simple bubble summation involves a summation of all bubble diagrams to all 
orders. The dressed propagator is given by 
\bea
\Delta_d(p)&=&\frac{1}{p^2+m^2+\Sigma(p)}.
\eea
The dressed mass $M$ is determined from the self energy using
\bea
M=\sqrt{m^2+\Sigma(iM)}.
\label{mdre}
\eea
The self energy for the simple bubble sum (in 0+1d) is given by
\bea
 \Sigma(p)&=&-e^2\bigg\lmoustache_{-\infty}^{\infty}
  \frac{dk}{2\pi}\frac{1}{(k^2+\mu^2)}
  \nonumber\\
  &&\times\bigg\{\frac{(2p-k)^2}{[(p-k)^2+m^2]}-1\bigg\}.
\eea
The self energy integral in this case is trivial and can be performed 
analytically, and the dressed mass is determined from Eq.~(\ref{mdre})

\begin{figure}
\setcaptionmargin{5mm}
%\onelinecaptionsfalse
\includegraphics[width=3.0in]{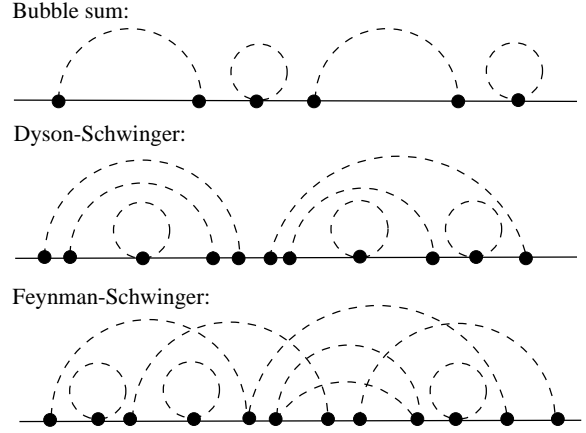}
%\includegraphics[width=3.0in]{allthree}
%captionstyle{normal}
\caption{ Various interactions included in each approach are shown.}
\label{allthree.fig} 
\end{figure}

The rainbow Dyson-Schwinger equation sums more diagrams than the simple 
bubble summation (Fig.~\ref{allthree.fig}). The self energy of the 
rainbow Dyson-Schwinger equation involves a momentum dependent mass:
\bea
\Sigma(p)&=&-e^2\bigg\lmoustache_{-\infty}^{\infty}
\frac{dk}{2\pi}\frac{1}{(k^2+\mu^2)}
\nonumber\\
&\times&\bigg\{\frac{(2p-k)^2}{[(p-k)^2+\underbrace{m^2+\Sigma(k)}]}-1\bigg\}\, .\qquad
\label{selfe}
\eea
In this case the self energy is nontrivial and it must be determined by
a numerical solution of Eq.~(\ref{selfe}). The dressed mass is determined 
by the logarithmic derivative of the dressed propagator in coordinate space  
\bea
M&=&-\lim_{T\rightarrow \infty}\frac{d}{dT}\,\, {\rm log}[\,\Delta_d(t)\,].
\eea

\begin{figure}
\setcaptionmargin{5mm}
\onelinecaptionsfalse
\includegraphics[width=3.5in]{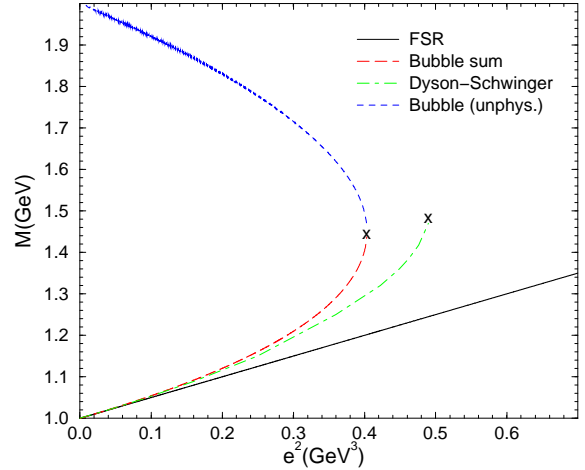}
%\includegraphics[width=3.5in]{mvsg2-sqed-collide}
%captionstyle{normal}
\caption{ The function $M(e^2)$, calculated in 0+1d with values of $m=\mu=1$ GeV, using the FSR approach, 
the Dyson-Schwinger equation in the rainbow approximation, 
and the bubble summation. 
While the exact result is always real, the rainbow DSE and the 
bubble summation results become complex beyond a critical coupling.}
\label{mvsg2.sqed.collide.fig} 
\end{figure}

The type of diagrams summed by each method is shown in Fig.~\ref{allthree.fig}.
Note that the matter loops do not give any contribution as explained earlier.
Results obtained by these three methods are shown in 
Fig.~\ref{mvsg2.sqed.collide.fig}.
It is interesting to note that the simple bubble summation and the rainbow
Dyson-Schwinger results display similar behavior. While the exact result 
provided by the FSR linearly increases for all coupling strengths, both the 
simple bubble summation  and the rainbow Dyson-Schwinger results come to a 
critical point beyond which solutions for the dressed masses become complex. {\em This 
example very clearly shows that conclusions about the mass poles of propagators
based on approximate methods such as the rainbow Dyson-Schwinger equation can 
be misleading}. 

\begin{figure}
\setcaptionmargin{5mm}
\onelinecaptionsfalse
\includegraphics[width=3.0in]{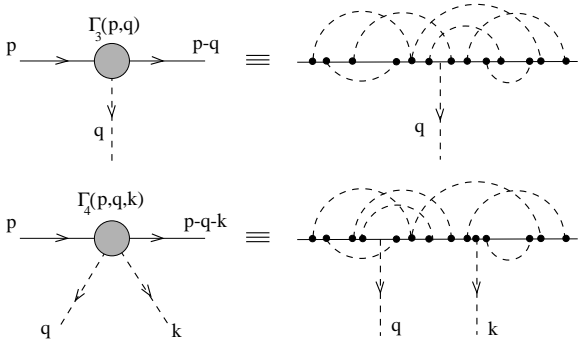}
%\includegraphics[width=3.0in]{definition}
%captionstyle{normal}
\caption{Because of the interactions, the one-particle
irreducable vertex functions $\Gamma_n$ $(n=3,4,\cdots)$ depend on the
external momenta.}
\label{definition} 
\end{figure}

\begin{figure}
\setcaptionmargin{5mm}
\onelinecaptionsfalse
\includegraphics[width=2.3in]{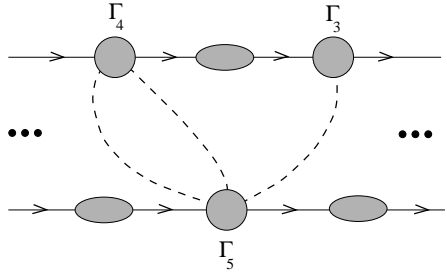}
%\includegraphics[width=2.3in]{2body}
%captionstyle{normal}
\caption{ Exact computation of the two-body bound state propagator requires
the summation of all particle self energies, vertex corrections, and ladder
and crossed ladder exchanges.}
\label{2body} 
\end{figure}

%\section{Vertex contributions in the FSR approach}
In general a consistent treatment of any nonperturbative calculation 
must involve summation of all possible vertex corrections. Vertex corrections
are those irreducible diagrams that surround an interaction vertex. 
%Take $\phi^3$ theory as an example. 
The elementary vertex is the
three-point  vertex, $\Gamma_3$, but the particle interactions will lead to
the appearance of $n$th order irreducible vertices, $\Gamma_n$, as
illustrated in Fig.~\ref{definition}.
%----------------------------------------------
%----------------------------------------------
The propagation of a bound state therefore involves a summation of all
diagrams with the inclusion of higher order vertices (Fig.~\ref{2body}). A
rigorous  determination of all of these vertices is not feasible. In the
literature on  bound states $\Gamma_{n>3}$ interaction vertices are usually
completely  ignored. The 3-point vertex $\Gamma_{3}$ can be approximately
calculated in  the ladder approximation~\cite{Maris:2000wz}. However a rigorous
determination of  the {\em exact} form of the 3-point vertex is not
possible, for this requires  the knowledge of even higher order vertices.
%----------------------------------------------

%----------------------------------------------

In order to be able to make a connection between the exact theory and
predictions based on approximate bound  state equations it is essential that
the role of interaction vertices be understood. The Feynman-Schwinger
Representation (FSR)  is a useful technique for this purpose. 
%In  this approach the path integrals over quantum  fields are integrated out and replaced by path integrals over  the trajectories of the particles. 

We now present the interesting outcome, that the full bound state
result dictated by a Lagrangian can be obtained by summing only generalized
ladder diagrams (``generalized'' ladders include both crossed ladders and, in
theories with an elementary four-point interaction, both overlapping and
non-overlapping ``triangle'' and ``bubble'' diagrams). 

\begin{figure}
\setcaptionmargin{5mm}
\onelinecaptionsfalse
\includegraphics[width=3.0in]{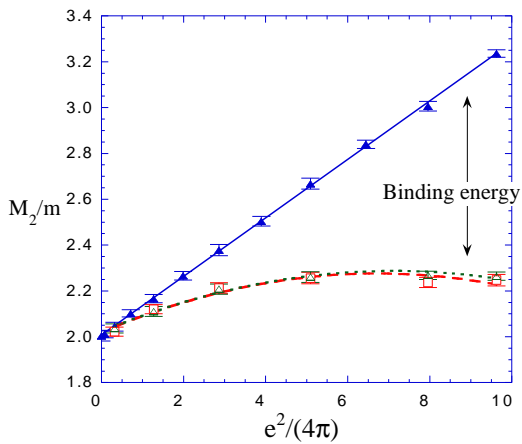}
%\includegraphics[width=3.0in]{M-e2-3+1a}
%captionstyle{normal}
\caption{Two-body bound state mass for SQED in 3+1 dimensions. Solid
triangles are 2$M_1$, open squares are $M_2^{\rm exch}$, and open triangles
are $M_2^{\rm tot}$.  Here $\mu/m=0.15$, $\Lambda_1/\mu=3$, and
$\Lambda_2/\mu=5$.  The smooth lines are fits to the ``data''.  Note that 
$M_2^{\rm exch}=M_2^{\rm tot}$ to within errors.}
\label{mvsg2.dressed.eps} 
\end{figure}

We adopt the following procedure for determining the contribution of vertex
corrections in 3+1 dimension. We start  with an initial bare mass $m$ and
calculate the full two-body bound state result with the inclusion of all
interactions: generalized ladders, self energies and vertex corrections.
Let us denote the result for the exact two-body bound state mass by $M_2^{\rm
tot}(e^2,m)$, since it will be a function of the coupling strength
$e$ and the bare input mass $m$, and the superscript ``tot'' implies that all
interactions are summed. Next we calculate the dressed one-body mass
$M_1(e^2,m)$. Then using the dressed mass value $M_1(e^2,m)$ we calculate the
bound state  mass $M_2^{\rm exch}(e^2,M_1)$ {\em by summing only the
generalized exchange interaction  contributions}. In this last calculation
we sum only exchange interactions (generalized ladders), but the
self energy is approximately taken into account since we use the (constant)
dressed one-body mass as input. However the vertex  corrections and
wavefunction renormalization are completely left out since we  use the
original vertex provided by the Lagrangian. In order to compare the full
result where all interactions have been summed with the result obtained by
two dressed particles interacting only by generalized ladder
exchanges we plot the bound state masses obtained by these methods.
Numerical results are presented in  Fig.~\ref{mvsg2.dressed.eps}.  This
result is qualitatively similar to that obtained analytically for SQED in 
0+1 dimension~\cite{Savkli:2002fj}.   
%
%----------------------------------------------
%----------------------------------------------
%
 
\begin{figure}
\setcaptionmargin{5mm}
\onelinecaptionsfalse
\includegraphics[width=3.0in]{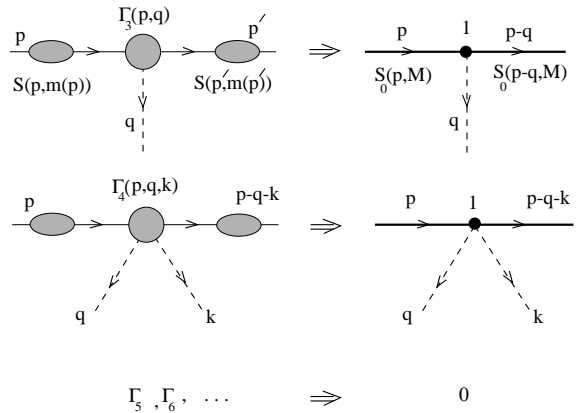}
%\includegraphics[width=3.0in]{equivalence}
%captionstyle{normal}
\caption{The correct two-body result can be obtained by simply using a
dressed constituent mass and a bare vertex, and ignoring the contributions
of higher order vertices.}
\label{mvsg2.sqed} 
\end{figure}

The numerical results presented here
yield the following prescription for bound state calculations: 
In order to get the full result for bound states it is 
a good approximation to first solve for dressed one-body masses exactly
(summing all generalized rainbow diagrams), and then use these
dressed masses and the  bare interaction vertex provided by the Lagrangian
to calculate the bound  state mass by summing only generalized ladder
interactions (leaving out  vertex corrections). In terms of Feynman graphs
this prescription can  be expressed as in Fig.~\ref{mvsg2.sqed}

%----------------------------------------------

%----------------------------------------------

The significance of the results presented above rests in the fact that the
problem of calculating exact results for bound state masses in SQED has been
reduced to that of calculating only generalized ladders. Summation of
generalized ladders can be addressed within the  context of bound state
equations~\cite{Gross:1969rv,Gross:1982nz,Phillips:1998uk}. Here we  have shown
the connection between the full prediction of a Lagrangian and the summation
of generalized ladder diagrams.  Our results are rigouous for
SQED, but are only suggestive for more general theories with spin or internal
symmetries.  Since we have neglected charged particle loops (our
results are in quenched approximation), and the current is conserved in
SQED, it is perhaps not surprising that the bare coupling is not
renormalized, but the fact that the momentum dependence of the dressed mass
and vertex corrections seem to cancel is unexpected.  If we were to
unquench our calculation, or to use a  theory without a conserved current,
it is reasonable to expect that {\it both\/} the bare interaction and the
mass would be renormalized. 

Finally, we call attention to a remarkable cancellation that occurs in the
one-body calculations.  The exact self energy shown in Fig.~\ref{mvsg2.dressed.eps}  (and also in Fig.~\ref{mvsg2.sqed.collide.fig} for different parameters)
is nearly linear in $e^2$
\cite{Savkli:1999ui}.  This remarkable fact implies that the {\it exact\/} self
energy is well approximated by the lowest order result from perturbation
theory.  It is instructive to see how this comes about.  If we expand the
self energy to fourth order, expanding each term about the bare mass $m$, we
have
\begin{eqnarray}
S^{-1}_d(p^2) &=&m^2-p^2 + \Sigma(p^2) \nonumber\\
    &=&m^2-p^2 + \Sigma_2 +(p^2-m^2)\Sigma_2' 
    \nonumber\\
    &&+\Sigma_4\, ,\quad
\label{self1}
\end{eqnarray}
where $\Sigma_\ell=\Sigma_\ell(m^2)$ is the contribution of order $e^\ell$
evaluated at $p^2=m^2$, $\Sigma'=d\Sigma(p^2)/dp^2$ evaluated at $p^2=m^2$, 
and the formula is valid for $p^2-m^2\simeq e^2$.  Expanding the dressed
mass in a power series in $e^2$
\begin{eqnarray}
M_1^2 &=&m^2 + m_2^2 +m_4^2 +\cdots \, , \label{mass1}
\end{eqnarray}
where $m_\ell^2$ is the contribution of order $e^\ell$, and substituting
into Eq.~(\ref{self1}), give
\begin{eqnarray}
M_1^2 &=&m^2 + \Sigma_2 +\Sigma_2'\Sigma_2+\Sigma_4 +\cdots \, ,
\label{mass2}
\end{eqnarray}
The mass is then
\begin{eqnarray}
M_1 &=&m + \frac{\Sigma_2}{2m}
+\frac{4m^2\left[\Sigma_2'\Sigma_2+\Sigma_4\right]-\Sigma_2^2}{8m^3} 
\nonumber\\
&&+\cdots
\, , \qquad \label{mass3} 
\end{eqnarray}
The linearity of the exact result implies that the forth order term in
Eq.~(\ref{mass3}) must be zero (or very small), and this can be easily
confirmed by direct calculations!  

The cancellation of the fourth order mass correction (and all higher orders)
is reminisent of the cancellations between generalized ladders that explains
why quasipotential equations are more effective that the ladder
Bethe-Salpeter equation in explaining 2-body binding energies.  It shows
that a simple evaluation of the second order self energy at the bare mass
point is more accurate than solution of the Dyson Schwinger equation in
rainbow approximation.

The general lesson seems to be that attempts to sum a small subclass of
diagrams exactly is often less accurate than the approximate summation of a
larger class of diagrams.
  
In the next section we consider the application of the FSR approach to 
scalar $\chi^2\phi$ interaction

\begin{figure}
\setcaptionmargin{5mm}
%\onelinecaptionsfalse
\includegraphics[width=3.0in]{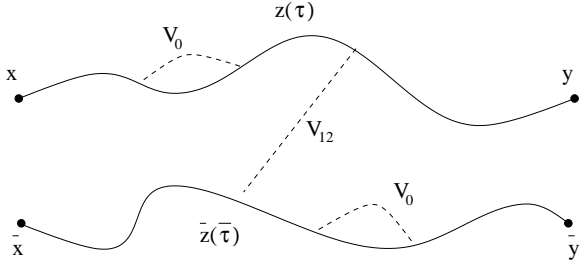}
%\includegraphics[width=3.0in]{trajectory}
%captionstyle{normal}
\caption{Sample trajectories with self and exchange interactions.}
\label{trajectory.fig} 
\end{figure}

\begin{figure}
\setcaptionmargin{5mm}
%\onelinecaptionsfalse
\includegraphics[width=2.2in]{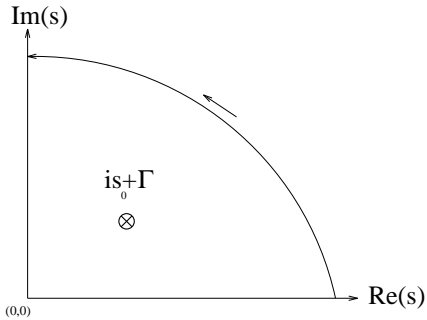}
%\includegraphics[width=2.2in]{wick}
%captionstyle{normal}
\caption{Wick rotation in the $s$ integration.}
\label{wick.fig} 
\end{figure}

\begin{figure}
\setcaptionmargin{5mm}
%\onelinecaptionsfalse
\includegraphics[width=3.0in]{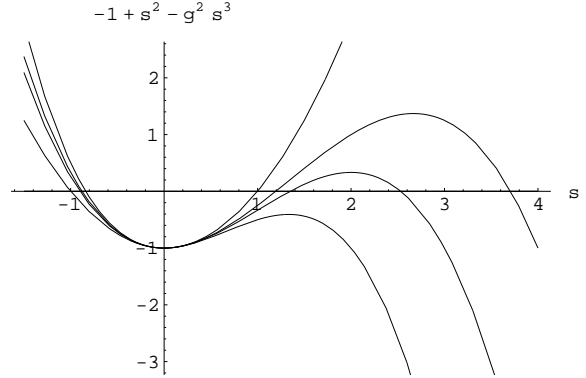}
%\includegraphics[width=3.0in]{gcritical}
%captionstyle{normal}
\caption{As the effective coupling strength g$^2$ is increased, the stationary point disappears.}
\label{gcritical.fig} 
\end{figure}

\begin{widetext}

\section{Scalar $\chi^2\phi$ interaction with the FSR approach}

We consider the theory of charged scalar particles $\chi$ of mass $m$ 
interacting through the exchange of a neutral scalar particle $\phi$ of mass 
$\mu$. The Euclidean Lagrangian for this theory is given by 
\bea
{\cal L}&=&\chi^*\bigl[m^2-\del^2+g\phi\bigr]\chi+\frac{1}{2}\,\phi(\mu^2-\del^2)\phi.
\eea
The 2-body propagator for the transition from the initial state
$\Phi_{i}=\chi^*(x)\chi(\bar{x})$ to final state $\Phi_{f}=\chi^*(y)\chi(\bar{y})$ is given by
\bea
&&G(y,\bar{y}|x,\bar{x})=N\bigg\lmoustache {\cal D}\chi^*\bigg\lmoustache {\cal D}\chi\bigg\lmoustache {\cal D}\phi\,\,\,\Phi^*_f\,\Phi_i\,\,{\rm exp}\biggl[-\int d^4x\,{\cal L}\biggr],
\eea
After the usual integration of matter fields is done, the Green's function 
reduces to
\bea
G(y,\bar{y}|x,\bar{x})&=&N\bigg\lmoustache {\cal D}\phi\, ({\rm det}S)\,\,S(x,y)S(\bar{x},\bar{y})\,e^{-L_0[\phi]}\, ,
\eea
with the free Lagrangian, $L_0$, and the interacting propagator, $S(x,y)$, defined by
\bea
L_0[\phi]&\equiv&\frac{1}{2}\int d^4 z \;\phi(z)(\mu^2-\partial_z^2)\phi(z) \nonumber\\
S(x,y)&\equiv &\langle y\,|\,\frac{1}{m^2+H(\hat{z},\hat{p})}\,|\,x\rangle
\, ,
\eea
with the Hamiltonian
\bea
H(\hat{z},\hat{p})&\equiv& \hat{p}^2-g\phi(\hat{z})\, .
\label{hphi3}
\eea
As in the case of scalar QED we employ the quenched approximation: ${\rm det} S\rightarrow 1$. 

We exponentiate the denominator by introducing an $s$ integration along the {\em imaginary} axis with an $\epsilon$ prescription 
\bea
S(x,y)&=&\int_0^{i\infty}ds\,\, \,e^{-s (m^2+i\epsilon)}\,\langle y\,|
\,{\rm exp}[-sH]\,|\,x\rangle \, .
\eea
This representation should be compared with the representation used earlier 
in SQED Eq.~(\ref{sqeds}). Here the integration is done along the imaginary axis
because $H$ is not positive definite.

Again, a quantum mechanical path integral 
representation can be constructed by recognizing that Lagrangian corresponding
to the $H$ of Eq.~(\ref{hphi3}) is given by 
\bea
L(z,\dot{z})=\frac{\dot{z}^2}{4}+g\phi(z)\, .
\eea
The path integral representation for the interacting propagator is therefore
\beas
S(x,y)&=&-i\bigg\lmoustache_0^\infty ds\,\bigg\lmoustache\,{\cal D}z\,\,{\rm exp}\biggl[is(m^2+i\epsilon)-\frac{i}{4}\int_0^s d\tau\, \dot{z}^2(\tau)+ig\int_0^s d\tau\,\phi(z(\tau))\,\,\biggr].
\eeas
This representation allows the elimination of the integral over the exchange 
field $\phi$. The 2-body propagator reduces to 
\bea
G&=&-\bigg\lmoustache_0^\infty ds\,\, \bigg\lmoustache_0^\infty d\bar{s}\,\, \bigg\lmoustache\, ({\cal D}z)_{xy}\,\bigg\lmoustache\, 
({\cal D}\bar{z})_{\bar{x}\bar{y}}\,\,e^{iK[z,s]+iK[\bar{z},\bar{s}]} I_\phi\, ,
\eea
where mass and kinetic term is given by
\bea
K[z,s]&=&(m^2+i\epsilon)s-\frac{1}{4s}\int_0^1 d\tau \,\dot{z}^2(\tau)\, .
\eea
The field integration $I_\phi$ is a standard gaussian integration
\bea
I_\phi &\equiv&\biggl\lmoustache 
{\cal D}\phi\,{\rm exp}\biggl[+ig\biggl( \int_0^s d\tau\,\phi(z(\tau)) +\int_0^{\bar{s}} d\bar{\tau}\,\phi(\bar{z}(\bar{\tau})) \biggr)  - L_0[\phi] \biggr]
\nonumber\\
&\equiv& {\rm exp}\biggl(-V_{0}[z,s]-2\,V_{12}[z,\bar{z},s,\bar{s}]-V_{0}[\bar{z},\bar{s}]\biggr),
\eea
where $V_{0}$ and $V_{12}$ (self and exchange energy contributions in 
Fig.~\ref{trajectory.fig}) are defined by
\bea
V_{0}[z,s]&=&\frac{g^2}{2}\,\,s^2\,\,\bigg\lmoustache_0^1d\tau\bigg\lmoustache_0^1d\tau' \,\Delta(z(\tau)-z(\tau'),\mu)
\equiv s^2\, v[z]\label{v0.eq}\\
V_{12}[z,\bar{z},s,\bar{s}]&=&\frac{g^2}{2}\,\,s\bar{s}\,\,\bigg\lmoustache_0^1d\tau\bigg\lmoustache_0^1d\bar{\tau}\, \Delta(z(\tau)-\bar{z}(\bar{\tau}),\mu)\, .
\eea
%
%\end{widetext}
It should be noted that the interaction terms explicitly depend on the $s$ 
variable, which was not the case for SQED. The interaction kernel $\Delta$ is 
given by
\bea
\Delta(x,\mu)&=& \bigg\lmoustache \frac{d^4p}{(2\pi)^4}\frac{e^{ip\cdot x}}{p^2+\mu^2}
\nonumber\\
&=&\frac{\mu}{4\pi^2|x|}K_1(\mu|x|)\, .
\eea

In order to be able to compute the path integral over trajectories,  a 
discretization of the path integral is needed
\bea
({\cal D}z)_{xy}&\rightarrow&(N/4\pi s)^{2N}\Pi^{N-1}_{i=1}\bigg\lmoustache d^4z_i\, .
\label{discreet}
\eea
The $s$ dependence is {\em crucial} for correct normalization. After 
%\begin{widetext}
discretization, the 1-body propagator takes the following form
\bea
G&=&i\left(\frac{N}{4\pi}\right)^{2N}\bigg\lmoustache \Pi_{i=1}^{N-1}dz_i \bigg\lmoustache_0^\infty \frac{ds}{s^{2N}}\,{\rm exp}\biggl[im^2s-i\frac{k^2}{4s}-s^2\,v[z]\biggr],
\eea
where $v[z]$ was defined in Eq.~(\ref{v0.eq}).
This is an {\em oscillatory} and {\em regular} integral and it is not 
convenient for Monte-Carlo integration. The origin of the oscillation is the 
fact that $s$ integral was defined along the imaginary axis,
\bea
&&{\rm Rep.\,\, 1:}\,\,\,\,S(x,y)=<y\,|\bigg\lmoustache_0^{-i\infty}ds\, {\rm exp}\biggl[-s(m^2-\del^2+g\phi+i\epsilon)\biggr]|\,x>\, .
\label{phi3.int.prop.2}
\eea
In earlier works~\cite{Simonov:1989mj,Nieuwenhuis:1996mc} a nonoscillatory Feynman-Schwinger 
representation was used,
\bea
&&\hspace{-1.2cm}
{\rm Rep.\,\, 2:}\,\,\,\,S(x,y)=<y\,|\bigg\lmoustache_0^\infty ds\, {\rm exp}\biggl[-s(m^2-\del^2+g\phi)\biggr]|\,x>\, .
\eea
Rep. 2 leads to a {\em nonoscillatory} and {\em divergent} result
\bea
G&\propto&\bigg\lmoustache_0^\infty \frac{ds}{s^{2N}}\,{\rm exp}\biggl[-m^2s-\frac{k^2}{4s}+s^2v[z]\biggr],
\eea
and the large $s$ divergence was regulated by a cut-off $\Lambda$. This is not
a satisfactory prescription since it relies on an arbitrary cut-off. Later 
it was shown~\cite{Savkli:1999rw,Savkli:1999gq} that the correct procedure is to start with
Rep.1 and make a Wick rotation such that the final result is {\em nonoscillatory}
and {\em regular}.
The implementation of Wick rotation however is nontrivial. Consider
the s-dependent part of the integral for the 1-body propagator
\bea
G&\propto&\bigg\lmoustache_0^\infty \frac{ds}{s^{2N}}\,{\rm exp}\biggl[im^2s-i\frac{k^2}{4s}-s^2\,v[z]\biggr],
\label{gsimple}
\eea
\end{widetext}
It is clear that a replacement of $s\rightarrow is$ leads to a divergent 
result. The problem with the Wick rotation (Fig~\ref{wick.fig}) comes from the 
fact that the $s$ integral is infinite both along the imaginary axis {\em and}
along the contour at infinity. These two infinities cancel to yield a finite 
integral along the real axis.
As $g\rightarrow 0$ the dominant contribution to the $s$ integral in 
Eq.~(\ref{gsimple}) comes from the stationary point 
\bea
s=is_0\simeq i\frac{k}{2m}.
\eea
Therefore one might suppress the integrand away from the stationary 
point by introducing a damping factor $R$
\bea
R(s,s_0)&\equiv& 1-(s-is_0)^2/\Gamma^2.
\eea
With this factor the integral in Eq.~(\ref{gsimple}) is modified 
\bea
G\propto\bigg\lmoustache_0^\infty\!\! \frac{ds}{s^{2N}}\,{\rm exp}\biggl[im^2s-i\frac{k^2}{4s}-\frac{s^2\,v[z]}{R^2(s,s_0)}\biggr]\, .\nonumber\\
\label{gcomplex}
\eea
This modification allows us to make a Wick rotation since the contribution 
of the contour at infinity now vanishes. However this procedure relies on the 
fact that there exists a stationary point. It can be seen from the original 
expression Eq.~(\ref{gsimple}) that this is not always true. According to the 
original integral the stationary point is given by the following equation
\bea
im^2+i\frac{k^2}{4s^2}-2s\,v[z]&=&0\, .
\label{ocrit}
\eea
Introducing $s=i{\rm s}\frac{k}{2m}$ and ${\rm g}^2=k\left<v[z]\right> /m^3$
this equation becomes
\bea
-1+{\rm s}^2-{\rm g}^2{\rm s}^3=0\, .
\eea
The stationary point is determined by the first intersection of a cubic plot 
with the positive s axis as shown in Fig.~\ref{gcritical.fig}.
The plot in Fig.~\ref{gcritical.fig} shows that, as the effective coupling strength g$^2$ is 
increased the curve no longer crosses the positive s axis. Therefore beyond 
a critical coupling strength the stationary point vanishes and mass results 
should be unstable.
Limiting discussion to cases where the original expression Eq.~(\ref{gsimple}) has a critical 
point, we now turn to perform a Wick rotation on the modified expression 
Eq.~(\ref{gcomplex}). The Wick rotation in Eq.~(\ref{gcomplex}) amounts to 
a simple replacement $s\rightarrow is$, and a {\it regular}, {\it nonoscillatory} integral is found:
\bea
G\propto\bigg\lmoustache_0^\infty \frac{ds}{s^{2N}}\,{\rm exp}\biggl[-m^2s-\frac{k^2}{4s}+\frac{s^2\,v[z]}{R^2(is,s_0)}\biggr]\, .
\nonumber\\
\label{Gone.eq}
\eea
At first look it seems that the new integral always has a stationary point
determined by the following equation
\bea
&&-m^2+\frac{k^2}{4s^2}+\frac{2s\,v[z]}{R^2(is,s_0)}
\nonumber\\
&&\qquad-s^2v[z]
\frac{(R^2(is,s_0))'}{R^2(is,s_0)}=0.
\label{ncrit}
\eea
The key point to remember is that the stationary point we find {\em after}
the Wick rotation should be the same stationary point we had before 
the Wick rotation. This is required to make sure that the physics remains 
the same after the Wick rotation. Self consistency therefore requires that 
the stationary point after the Wick rotation is at $s=is_0$. In that case 
$R(is_0,s_0)=1$, and $(R^2(is_0,s_0))'=0$ and Eq.~(\ref{ncrit})
determining the critical point  reduces to the earlier original
 Eq.~(\ref{ocrit}). 

The regularization of the ultraviolet singularities is done using 
Pauli-Villars regularization, which is particularly convenient for numerical integration since it only involves 
a change in the interaction kernel
\bea
\Delta(x,\mu)&\longrightarrow& \Delta(x,\mu)-\Delta(x,\alpha\mu).
\eea
%
%\section{Numerical applications of $\chi^2\phi$ interaction}

Calculations of the $\chi^2\phi$ interaction in 3+1d require numerical 
Monte-Carlo integration. The first step is to represent the particle trajectories by a discrete number of $N+1$ points with 
\begin{figure}
\setcaptionmargin{5mm}
\onelinecaptionsfalse
\includegraphics[width=2.2in]{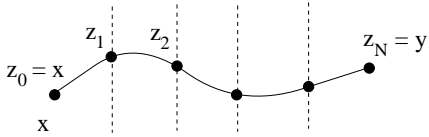}
%\includegraphics[width=2.2in]{discrete}
%captionstyle{normal}
\caption{Number of steps a particle takes between initial and 
final coordinates is discretized. The space-time is {\em continuous} and there 
are {\em no space-time boundaries}}
\label{discrete.fig} 
\end{figure}
boundary conditions given by
\bea
&&z_0=x=(x_1,x_2,x_3,0)
\nonumber\\
&&z_N=y=(y_1,y_2,y_3,T)\, .
\eea
The discretization employed in the FSR is for the {\em number of time steps} a particle
takes in going from the initial time to the final time along a trajectory in a 4-d coordinate space. This is 
very different from the discretization employed in lattice gauge theory. 
Contrary to lattice gauge theory, in the FSR approach space-time is 
{\em continuous} and rotational symmetry is respected. An additional, important 
benefit is the {\em lack of space-time} lattice boundaries, which allows calculations of arbitrarily large systems using the FSR approach. This feature
provides an opportunity for doing complex applications such as calculation of the form factors of large systems. 

In doing Monte-Carlo sampling we sample {\em trajectories} (lines) rather than
 {\em gauge field configurations} (in a volume). This leads to a significant 
reduction in the numerical cost. 
The ground state mass of the Green's function is obtained using
\begin{equation}
m_0=\frac{\int {\cal D}Z\, S'[Z]e^{-S[Z]}}{\int {\cal D}Z\, e^{-S[Z]}}.
\label{groundstate2}
\end{equation}
Sampling of trajectories is done using the standard Metropolis algorithm, which insures that configurations sampled are distributed 
according to the weight $e^{-S[Z]}$. In sampling trajectories the final state
(spacial) coordinates of particles can be integrated over, which puts the system at rest and  projects out the $S$-wave ground 
state. As trajectories of particles are sampled, the wave function of the system 
can be determined simply by storing the final state configurations of 
particles in a histogram. 

In sampling trajectories the first step is thermalization. In order to insure 
that the initial configuration of trajectories has no effect on results, the first 
1000 or so updates are not taken into account. 
Statistical independence of subsequent samplings is measured by the 
correlation function $X(n)$, defined as
\bea
X(n)&\equiv&\frac{\left< m(i) m(i+n) \right> - \left< m \right>^2 }{\left< m 
\right>^ 2 },
\eea
where $m(i)$ is the mass measurement at the $i$'th update. 

\begin{figure}
\setcaptionmargin{5mm}
\onelinecaptionsfalse
\includegraphics[width=3.0in]{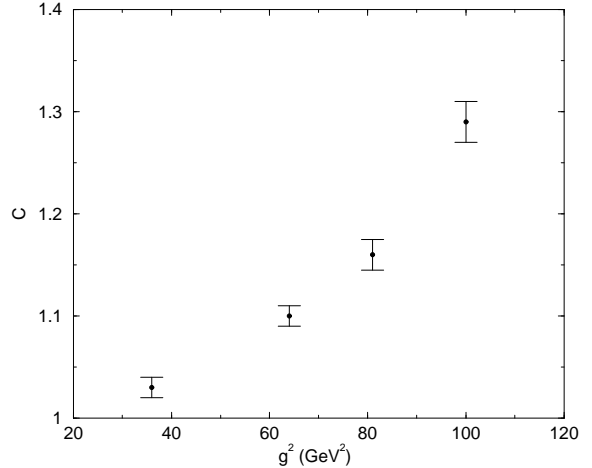}
%\includegraphics[width=3.0in]{cvsg2-2b}
%captionstyle{normal}
\caption{The dependence of the peak of s-distribution on the coupling
strength for the 2-body bound state is shown. The peak location is given by $s_0=C T/2m$. There is no real
solution for $C$ beyond $g^2=100$ GeV$^2$.}
\label{cvsg2.2b.fig} 
\end{figure}

In order to insure that the location of the stationary point is self consistent, as discussed earlier, its location must be 
determined carefully. The stationary point can be parametrized by 
$s_0=CT/(2m)$, where $T/2m$ is the location of the stationary point 
when the coupling strength $g$ goes to zero. As the coupling strength
is increased, the stationary point moves to larger values of $s_0$ (recall Fig.~\ref{gcritical.fig}) and $C$ increases. 
Eventually a critical value of the coupling constant  is reached beyond which there is no self consistent stationary point. In order to be able to do Monte-Carlo integrations, an initial
guess must be made for the location of the stationary point. Self consistency 
is realized by insuring that the peak location of the s distribution in the 
Monte-Carlo integration agrees with the initial guess for the stationary 
point~\cite{Savkli:1999gq}. In Fig.~\ref{cvsg2.2b.fig} the dependence of the location of 
the stationary point on the coupling strength for 2-body bound states is shown. The figure
shows that beyond the critical point $g^2\simeq 100$ GeV$^2$, $C$ goes to 
infinity implying that there is no stationary point. A similar critical 
behavior was also observed in Refs.~\cite{Rosenfelder:1996bd,Rosenfelder:1996ra} within the context of a variational approach.
\begin{figure}
\setcaptionmargin{5mm}
\onelinecaptionsfalse
\includegraphics[width=3.0in]{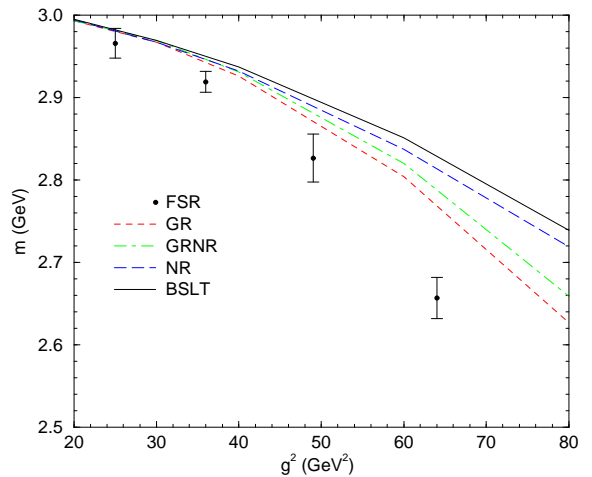}
%\includegraphics[width=3.0in]{mvsg2-phi3-2b}
%captionstyle{normal}
\caption{The coupling constant dependence of the 2-body bound state mass is shown. 
Beyond the critical coupling strength of $g^2=100$ GeV$^2$ 
the 2-body mass becomes unstable. The Bethe-Salpeter equation 
in the ladder approximation gives the smallest binding.  The other models are described in the text. }
\label{mvsg2.phi3.2b.fig} 
\end{figure}
In Fig.~\ref{mvsg2.phi3.2b.fig} FSR 2-body bound state mass
results are shown  for $m_\chi=1$ GeV, $\mu_\phi=0.15$ GeV. 
These results are all for a Pauli-Villars mass of 3$\mu$.
Also are shown the predictions of various integral equation
calculations. 

The FSR calculation sums all ladder and crossed ladder diagrams, and excludes the self energy contributions.
According to Fig.~\ref{mvsg2.phi3.2b.fig} all bound state equations underbind. Among the manifestly covariant equations the 
Gross equation (labeled GR in Fig.~\ref{mvsg2.phi3.2b.fig}) gives the closest result to the exact calculation obtained by 
the FSR method. This is due to the fact that in the limit of infinitely 
heavy-light systems the Gross equation effectively sums all ladder and 
crossed ladder diagrams. The equal-time equation (labeled ET in Fig.~\ref{mvsg2.phi3.2b.fig}) also produces a strong binding, 
but the inclusion of retardation effects pushes the ET results away 
from the exact results (Mandelzweig-Wallace equation~\cite{Wallace:1989nm}, labeled MW in Fig.~\ref{mvsg2.phi3.2b.fig}). In 
particular the 
Bethe-Salpeter equation in the ladder approximation (labeled BSE in 
Fig.~\ref{mvsg2.phi3.2b.fig}) gives the lowest binding. Similarly the 
Blankenbecler-Sugar-Logunov-Tavkhelidze equation~\cite{Logunov:1963yc,Blankenbecler:1966gx} (labeled BSLT in Fig.~\ref{mvsg2.phi3.2b.fig}) gives a 
very low binding. A comparison of the 
ladder Bethe-Salpeter, Gross, and the FSR results shows that {\em the exchange
of crossed ladder diagrams plays a crucial role.}

\begin{figure}
\setcaptionmargin{5mm}
\onelinecaptionsfalse
\includegraphics[width=3.0in]{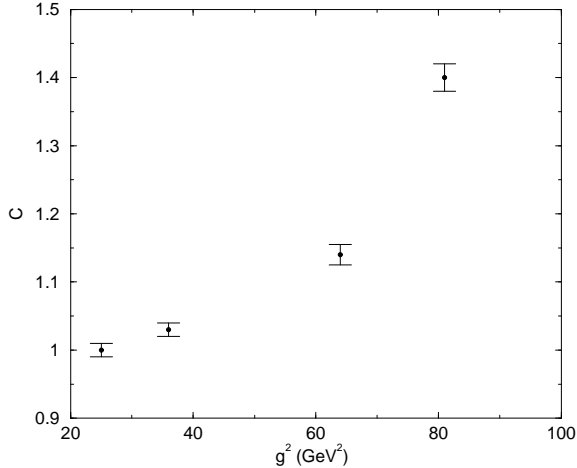}
%\includegraphics[width=3.0in]{cvsg2-3b}
%captionstyle{normal}
\caption{The dependence of the peak of s-distribution on the coupling
strength for the 3-body bound state mass is shown. The peak location is given
by $s_0=C T/2m$. There is no real solution for $C$ beyond $g^2=81$ GeV$^2$.}
\label{cvsg2.3b.fig} 
\end{figure}

\begin{figure}
\setcaptionmargin{5mm}
%\onelinecaptionsfalse
\includegraphics[width=3.6in]{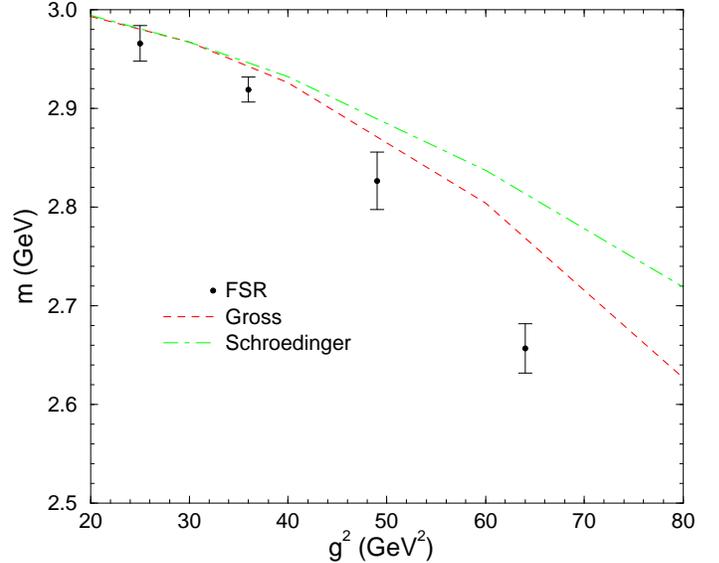}
%\includegraphics[width=3.6in]{mvsg2-phi3-3b-col}
%captionstyle{normal}
\caption{3-body bound state results for 3 equal 
mass particles of mass 1 GeV.}
\label{mvsg2.phi3.3b.fig} 
\end{figure}

In Fig.~\ref{cvsg2.3b.fig} the dependence of the location 
of  the stationary point on the coupling strength for 3-body bound states is shown.
In Fig.~\ref{mvsg2.phi3.3b.fig} the 3-body bound state results for 3 equal 
mass particles of mass 1 GeV are shown. For the 3-body case the only available 
results are for the Schr\"odinger and Gross equations. According to the results
presented in Fig.~\ref{mvsg2.phi3.3b.fig}, the bound state equations underbind 
for the 3-body case too. The Gross equation gives the closest result to the exact 
FSR result.

\begin{figure*}
\setcaptionmargin{5mm}
\onelinecaptionsfalse
\includegraphics[width=5.0in]{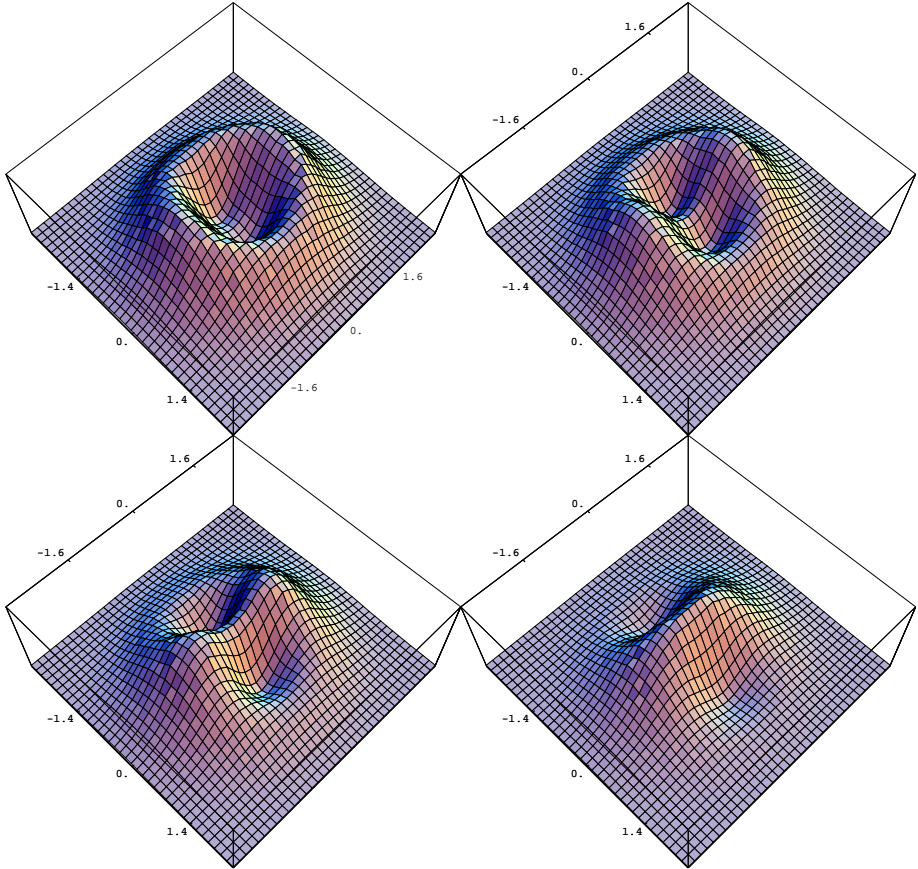}
%\includegraphics[width=5.0in]{surface-3b}
%captionstyle{normal}
\caption{The four panels show the evolution of the probability distribution for the 3rd particle as the distance between the two fixed particles is increased. When the fixed particles are very close to each other the third particle sees them as a single particle (the upper left plot). As the fixed particles are separated, the third particle starts penetrating between them (2nd and 3rd panels), and when two fixed particles are far apart (as shown in the lower right panel), the third particle is most likely to be found between the two fixed particles. }
\label{rdist.surf.3b} 
\end{figure*}

Determination of the wavefunction of bound states is done by keeping the final 
state configurations of particles in a histogram. For example, for a 3-body 
bound state system, the probability distribution of the third particle for a 
given configuration of first and second particles is shown in 
Fig.~\ref{rdist.surf.3b}. In the upper-left panel of 
Fig.~\ref{rdist.surf.3b}, the two fixed particles are very close to each other so
that the third particle sees them as a single particle. However as the fixed 
particles are separated from each other the third particle starts having 
a nonzero probability of being in between the two fixed particles (separation increases as we go from upper right to lower left panels in Fig.~\ref{rdist.surf.3b}). Eventually when the 
two fixed particles are far away from each other the third particle has 
a nonzero probability distribution only at the origin (the lower left panel in Fig.~\ref{rdist.surf.3b}).

Up to this point the FSR method has been derived and various applications 
to nonperturbative problems have been presented. In the next section we discuss
the stability of the $\chi^2\phi$ theory.

\section{Stability of the scalar $\chi^2\phi$ interaction}

Scalar field theories with a $\chi^\dag\chi\phi$ interaction (which we 
will subsequently denote simply by $\chi^2\phi$)  have been used
frequently without any sign of instability, despite an argument in 1952 by
Dyson\cite{Dyson:1952tj} suggesting instability, and a proof in 1959 by 
G.~Baym~\cite{Baym} showing that the theory is unstable.  For example, it is easy
to show that, for a limited range of coupling values $0\le g^2\le g^2_{crit}$, the
simple sum of bubble diagrams for the propagation of a single $\chi$ particle
leads to a stable ground state, and it is shown in Ref.~\cite{Savkli:1999me} that a
similar result also holds for the {\it exact\/} result in ``quenched''
approximation.  However, if the scalar
$\chi^2\phi$ interaction is unstable, then this instability should be
observed even when the coupling strength $g$ is vanishingly small
$g^2\rightarrow 0^+$, as pointed out recently by Rosenfelder and
Schreiber\cite{Rosenfelder:1996bd} (see also Ref.~\cite{Ding:1999ed}).  Both the simple bubble summation
and the quenched  calculations do not exhibit this behavior.  Why do the simple 
bubble  summation and the exact quenched calculations produce stable results for a
finite range of coupling values?

A clue to the answer is already provided by the simplest
semiclasical estimate of the ground state energy.  In this  approximation the
gound state energy is obtained by minimizing 
\be
E_0=m^2\chi^2 + \frac{1}{2}\mu^2 \phi^2 -g\phi\chi^2\, ,
\ee
where $m$ is the bare mass of the matter particles, and $\mu$ the mass
of the ``exchanged'' quanta, which we will refer to as the {\it
mesons\/}.  The minimum occurs at 
\be
E_0=m^2\chi^2 - g^2\frac{\chi^4}{2\mu^2}\, .
\ee
This is identical to a $\chi^4$ theory with a coupling of the wrong
sign, as discussed by Zinn-Justin\cite{Zinn-Justin:1989mi}.  The ground state is therefore
stable (i.e. greater than zero) provided
\be
g^2<g^2_{\rm crit}=\frac{2m^2\mu^2}{\chi^2}\, . 
\ee
This simple estimate suggests that the theory is stable over a
limited range of couplings {\it if the strength of the $\chi$ field is
finite\/}.  We now develop this argument more precisely and show
under what conditions it holds.

Before presenting new results, we lay the foundation using the variational
principle.  In the Heisenberg representation the fields are expanded in terms of
creation and annihilation operators that depend on time
\bea
\chi(t,\bf{r})&=&\int d\tilde{k}_m \left[
a(k)\,e^{-ik\cdot x} + b^{\dag}(k)
\,e^{ik\cdot x}\right]  \nonumber\\
\phi(t,\bf{r})&=&\int d\tilde{k}_\mu \left[
c(k)\,e^{-ik\cdot x} + c^{\dag}(k)
\,e^{ik\cdot x}\right]\, , \label{fields}
\nonumber\\
\eea
where $x=\{t, {\bf r}\}$ and 
\be
d\tilde{k}_m \equiv \frac{ d^3k } { (2\pi)^3\,2E_m(k) }
\ee
with $E_m(k)=\sqrt{m^2+k^2}$. The equal-time commutation relations are 
\be
\left[a(k),a^\dag(k')\right] = (2\pi)^3\,2E_m(k)\, \delta^3(k-k')\, .
\ee

The Lagrangian for the $\chi^2\phi$ theory is  
\begin{widetext}  
\be
{\cal L}=\chi^\dag\left[\del^2-m^2+g\phi\right]\chi+\frac{1}{2}\,
\phi\left(\del^2-\mu^2\right)\phi\, ,
\label{lagr0}
\ee
and the Hamiltonian $H$ is a normal ordered product of interacting (or dressed)
fields $\phi_d$ and $\chi_d$
\bea
H[\phi_d,\chi_d,t\,] 
= \int d^3r\, : \Biggl\{
\left(\frac{\del\chi_d}{\del t}\right)^2+({\bf\nabla}\chi_d)^2+m^2 \chi_d^2
\nonumber\\ 
+\frac{1}{2}\biggl[ \left(\frac{ \del \phi_d }{\del t}
\right)^2+(\vec{\nabla}\phi_d)^2+\mu^2\phi_d^2 \biggr] 
- g \chi_d^2 \phi_d \Biggr\}:\, . \label{hamiltonian}
\eea
This hamitonian conserves the {\it difference\/} between 
number of matter and the number of antimatter
particles, which we denote by $n_0$.  Eigenstates of the Hamiltonian
will therefore be denoted by
$\left|n_0,\lambda\right>$, where $\lambda$ represents the other quantum
numbers that define the state. Hence, allowing for the fact that the eigenvalue
may depend on the time,
\bea
 H[\phi_d,\chi_d,t\,] \left|n_0,\lambda\right> = M_{n_0,\lambda}(t) 
\left|n_0,\lambda\right>\, . \label{eveqn}
\eea

In the absence
of an exact solution of (\ref{eveqn}), we may estimate it from the
equation
\bea
M_{n_0,\lambda}(t)=&&\left<n_0,\lambda\right| H[\phi_d,\chi_d,t\,]
\left|n_0,\lambda\right>\nonumber\\
=&& \left<n_0,\lambda\right|U^{-1}(t,0) H[\phi,\chi,0\,]U(t,0)
\left|n_0,\lambda\right>\nonumber\\
\equiv&& \left<n_0,\lambda, t\right| H[\phi,\chi,0\,]
\left|n_0,\lambda,t\right> \, ,\label{matrixelem}
\eea
where $U(t,0)$ is the time translation operator which carries the Hamiltonian from
time $t=0$ to later time $t$.  We have also chosen $t=0$ to be the time at which
the interaction is turned on, $\phi_d(t)=U^{-1}(t,0)\phi(0)U(t,0)$, and the last
step simplifies the discussion by permitting us to work with a Hamiltonian
constructed from the {\it free\/} fields $\phi$ and $\chi$.  [If the interaction
were turned on at some other time
$t_0$, we would obtain the same result by absorbing the additional phases
$\exp(\pm i Et_0)$ into the creation and annhilation operators.]         

At $t=0$ the Hamiltonian in normal order reduces to
\bea
H[\phi,\chi,0\,] &=&\int d\tilde{k}_m\,E_m(k)\,{\cal N}_0(k,k)
+\int d\tilde{p}_\mu\,E_\mu(p)\, c^\dag(p)c(p) \nonumber\\ 
&&-\frac{g}{2}\int\,\frac{d\tilde{k}_m\,d\tilde{k'}_m}
{\omega(k-k')}{\cal N}_1(k,k')\biggl[c^\dag(k'-k) +c(k-k')\biggr],
\eea
where 
\bea
&&{\cal N}_0(k,k')=\left\{a^\dag(k)a(k')+b^\dag(k)b(k')\right\},\nonumber\\
&&{\cal N}_1(k,k')={\cal N}_0(k,k')+\left\{a^\dag(k)b^\dag(-k')+a(-k)b(k')\right\} 
\eea
and
$\omega(k)=\sqrt{\mu^2+{\bf k}^2}$.   To evaluate the matrix element
(\ref{matrixelem}) we express the the eigenstates as a sum of free particle states
with $n_0$ matter particles, $n_{\rm pair}$ pairs of $\chi{\bar\chi}$ particles,
and
$\ell$ mesons:
\bea
\left|n_0,\lambda,t\right>\equiv
\left|n_0,\alpha(t),\beta(t)\}\right>
=\frac{1}{\gamma(t)}\,\sum_{n_{\rm pair}=0}^\infty
\sum_{\ell=0}^\infty\alpha_{n_{\rm pair}}(t)
\beta_\ell(t)\left|n_0,n_{\rm pair},\ell\right>, \label{general}
\eea 
where $\gamma(t)$ is a normalization constant (defined below), the time
dependence of the states is contained in the time dependence of the
coefficients $\alpha(t)$ and $\beta(t)$,  and
\bea
\left| n_{0},n_{\rm pair},\ell\right>\equiv \int
\frac{\left|k_1,\cdots,k_{n_1};q_1,\cdots,q_{n_2};p_1,\cdots,p_\ell\right>}
{\sqrt{(n_0+n_{\rm pair})!\,n_{\rm pair}!\,\ell!}}
\label{fockstate}
\eea
with $n_1=n_0+n_{\rm pair}$, $n_2=n_{\rm pair}$ and
\be
\int=\int\,\prod_{i=1}^{n_1}
\,d\tilde{k_{i}}\,f(k_i)\;\prod_{j=1}^{n_2}\,
d\tilde{q_j}\,f(q_j)\;\prod_{l=1}^{\ell}\,
d\tilde{p_l}\,g(p_l) . 
\ee
The particle masses in $d\tilde{k}$ and $d\tilde{p}$ have been
suppressed; their values should  be clear from the context.
The normalization of the functions $f(p)$ and $g(p)$ is chosen to be
\be
\int d\tilde{k} \, f^2(k)=\int d\tilde{p}\, g^2(p)\equiv 1
\ee
which leads to the normalization 
\bea
\left<n'_{0},n'_{\rm pair},\ell'\,|\,n_{0},n_{\rm pair},\ell
\right>=&&\delta_{n'_{0},n_{0}}\,\delta_{n'_{\rm pair},n_{\rm
pair}}\, \delta_{\ell',\ell} \nonumber\\
\left<n_{0},\lambda,t\,|\,n_{0},\lambda,t
\right>=&&1  \, ,
\eea
if $\gamma(t)=\alpha(t)\beta(t)$ with
\bea
\alpha^2(t)=&&\sum_{n_{\rm pair}=0}^{\infty} \alpha_{n_{\rm
pair}}^2(t) = \alpha(t)\cdot\alpha(t)  \nonumber\\
\beta^2(t)=&&\sum_{\ell=0}^{\infty}\,\beta_{\ell}^2(t) =
\beta(t)\cdot\beta(t)
\, .
\eea
The expansion coefficients
$\{\alpha_{n_{\rm pair}}(t) \}$ and $\{\beta_\ell(t)\}$ are vectors in
infinite dimensional spaces.  

In principle the scalar cubic interaction in four dimensions requires 
ultraviolet regularization. However the issue of regularization and 
the question of stabilty are qualitatively unrelated. For example, the 
cubic interaction is also unstable in dimensions lower than four, where 
there is  no need for regularization. The ultraviolet regularization
would  have an effect on the behavior of functions $f(p)$, and $g(p)$,
which  are left  unspecified in this discussion except for their
normalization.

The matrix element (\ref{matrixelem}) can now be evaluated.  Assuming that
$f(k)=f(-k)$ and $g(k)=g(-k)$, it becomes:
\bea
M_{n_0,\lambda}(t)
=\left\{n_0+2L(t)\right\}\tilde{m} +G(t)\,\tilde{\mu}
-g V \left\{n_0+2L(t) +2L_1(t)\right\}\,
\sqrt{G_1(t)} \, , \label{generalelement}
\eea
where the constants $\tilde{m}$, $\tilde{\mu}$, and $V$ are
\bea
\tilde{m}&\equiv&\int\,d\tilde{k}\,E_m(k)\,f^2(k)\, ,\quad
\tilde{\mu}\equiv\int\,d\tilde{p}\,E_\mu(p)\,g^2(p)\nonumber\\
V&\equiv&\int\,\frac{d\tilde{k}_m\,d\tilde{k}'_m\,
f(k)f(k')g(k- k')} {\sqrt{m^2+({\bf k}-{\bf k}')^2}}\, ,
\eea
and the time dependent quantities are 
\bea
L(t)=&&\sum_{n_{\rm pair}=0}^\infty
\frac{n_{\rm pair}\,\alpha^2_{n_{\rm pair}}(t)}{\alpha^2(t)},\quad
G(t)=\sum_{\ell=0}^\infty
\frac{\ell\,\beta_\ell^2(t)}{\beta^2(t)}\nonumber\\ 
L_1(t)=&&\sum_{n_{\rm
pair}=1}^\infty
\frac{\sqrt{n_0+n_{\rm pair}}\sqrt{n_{\rm
pair}}\;\alpha_{n_{\rm pair}}(t)\,
\alpha_{{n_{\rm pair}}-1}(t)} {\alpha^2(t)}
\nonumber\\ 
\sqrt{G_1(t)}=&&\sum_{\ell=1}^\infty \frac{\sqrt{\ell}\;
\beta_\ell(t)\beta_{\ell-1}(t)}{\beta^2(t)} \, . \label{averages}
\eea
Note that $L$ and $G$ are the {\it average\/} number of matter pairs and mesons,
respectively, in the intermediate state.  

The variational principle tells us that the correct mass must be
equal to or larger than (\ref{generalelement}).  This inequality may be
simplified by using the Schwarz inequality to place an upper limit on the
quantities $L_1$ and $G_1$.  Introducing the vectors
\bea
f_1=&&\{\alpha_1,\sqrt{2}\,\alpha_2,\cdots\}=\{\sqrt{n}\;\alpha_{n}\}
\nonumber\\ 
f_2=&&\{\sqrt{n_0+1}\,\alpha_0,\sqrt{n_0+2}\,\alpha_1,\cdots \}
=\{\sqrt{n_0+n}\;\alpha_{n-1}\}\nonumber\\ 
h=&&\{\beta_1,\sqrt{2}\,\beta_2,\cdots
\} =\{\sqrt{\ell}\;\beta_{\ell}\}\, ,
\eea    
we may write 
\bea
L_1(t)=\frac{f_1(t)\cdot f_2(t)}{\alpha^2(t)}&&\le 
\frac{\sqrt{f_1^2(t)\,f_2^2(t)}}{\alpha^2(t)}\nonumber\\
&&=\sqrt{L(t)\{n_0+1+L(t)\}}\nonumber\\ 
\sqrt{G_1(t)}=\frac{h(t)\cdot \beta(t)}{\beta^2(t)}&&\le
\frac{\sqrt{h^2(t) \beta^2(t)}}{\beta^2(t)}=\sqrt{G(t)}\, . \label{conditions}
\eea
Hence, suppressing explicit reference to the time dependence of $L$ and $G$,
Eq.~(\ref{generalelement}) can be written
\bea
M_{n_0,\lambda}(t)
\ge\left(n_0+2L\right)\tilde{m} +G\,\tilde{\mu}
-g V
\left\{\left(\sqrt{n_0+1+L}+\sqrt{L}\right)^2-1\right\}\,
\sqrt{G} \, . \label{phivariation}
\eea
\end{widetext}
Minimization of the ground state energy with respect to the average number of
mesons $G$ occurs at
\be
\sqrt{G_0}=\frac{gV}{2\tilde{\mu}}\,\left\{\left(\sqrt{n_0+1+L}+
\sqrt{L}\right)^2-1\right\} \, .
\ee
At this minimum point the ground state energy is bounded by
\bea
M_{n_0,\lambda}(t)\geq&& \left\{n_0+2L\right\}\tilde{m}-\mu G_0\, .
\label{variational}
\eea
If we continue with the minimization process we would obtain
$M_{n_0,\lambda}(t)\to-\infty$ as $L\to\infty$, providing no lower bound and
hence suggesting that the state is unstable.  However, if {\it L is finite\/},
this result shows that the ground state is stable for couplings in the interval
$0<g^2<g_{crit}^2$ with
\be
g_{crit}^2\equiv\frac{4\,\tilde{\mu}\,\tilde{m}\,
(n_0+2L)}{V^2\left\{\left(\sqrt{n_0+1+L}+\sqrt{L}\right)^2-
1\right\}^2} \, .
\label{gcrit}
\ee
This interval is nonzero if the number of matter particles, $n_0$, and
the average number of $\chi\bar{\chi}$ pairs, $L$, is finite.  In
particular, {\it if there are no $Z$ diagrams or $\chi\bar{\chi}$ loops
in the intermediate states, then the ground state will be stable for a
limited range of values of the coupling.\/}

This result also suggests strongly that the system is unstable when
$g^2>g^2_{\rm crit}$, or when $L\to\infty$ (implying that $g^2_{\rm
crit}\to0$).  However, since Eq.~(\ref{variational}) is only a lower
bound, our argument does not provide a proof of these latter assertions. 

We now discuss the effect of the Z-graphs and of the matter loops on the stability. 
Using the Feynman-Schwinger representation (FSR), we will show that the
ground state is (i) {\it stable\/} when $Z$-diagrams are included in intermediate
states, but (ii)  {\it unstable\/} when matter loops are included.

The covariant trajectory $z(\tau)$ of the  particle
is parametrized in the FSR as a function of the proper time $\tau$. In $\chi^2\phi$
theory the FSR  expression for the 1-body propagator for a dressed
$\chi$-particle in quenched approximation in Euclidean space was given qualitatively in Eq.~(\ref{Gone.eq}).  The detailed expression, needed in the following discussion, is  
\begin{eqnarray}
G(x,y)&=&\int_0^{\infty} ds \left[\frac{N}{4\pi s}\right]^{2N}
\;\prod^{N-1}_{i=1}\int d^4z_i\nonumber\\
&&\times\exp\biggl\{-K[z,s]-V[z,s_r]\biggr\}\, ,\qquad \label{FS}
\end{eqnarray}
where the integrations are over all possible particle trajectories
(discretized into $N$ segments with $N-1$ variables $z_i$ and boundary
conditions $z_0=x$, and $z_N=y$) and the kinetic and self energy
terms are 
\bea
K[z,s]&=& m^2s+\frac{N}{4s}\sum_{i=1}^{N}(z_i-z_{i-1})^2\, ,\label{kin}\\
V[z,s]&=&-\frac{g^2s^2}{2N^2}\sum_{i,j=1}^{N}
\Delta\left(\delta z_{ij},\mu\right)\,  , 
\label{pot}
\eea
where $\Delta(z,\mu)$ is the Euclidean progagator of the meson
(suitably regularized), $\delta z_{ij}=\frac{1}{2}(z_i+z_{i-1}-z_j-z_{j-1})$, and  
\be
s_r\equiv \frac{s}{R(s,s_0)}=\frac{s}{1+(s-s_0)^2/\Gamma^2}\, .
\ee
(The need for the substitution $s\to s_r$ was discussed above in Sec.~IV.)      

In preparation for a discussion of the effects of $Z$-diagrams and
loops, we first discuss the stability of Eq.~(\ref{FS}) when neither $Z$-diagrams
nor loops are present.  To make the discussion explicit, consider the one
body propagator in 0+1 dimension.  Since the integrals converge, we make the
crude approximation that each $z_i$ integral is approximated by {\it one\/}
point (since we are excluding $Z$-diagrams, the points may lie along the classical
trajectory).  If the boundary conditions are
$z_0=0$ and
$z_N=T$ the points along the classical trajectory are $z_i=iT/N$, and 
\begin{equation}
K[z,s]= m^2s+\frac{N}{4s}\sum_{i=1}^{N}(z_i-z_{i-1})^2=m^2s+\frac{T^2}{4s}\, .
\label{KE}
\end{equation}
If the interaction is zero, this has a stationary point at $s=s_0=T/(2m)$,
giving 
\be
K[z,s]=K_0=mT\, ,
\ee 
yielding the expected free particle mass $m$.  [Note
that {\it half\/} of this result comes from the sum over
$(z_i-z_{i-1})^2$.]  The potential term (\ref{pot}) may be similarily
evaluated; it gives a negative contribution that reduces the mass.      

\begin{figure}
\setcaptionmargin{5mm}
%\onelinecaptionsfalse
\includegraphics[width=2.5in]{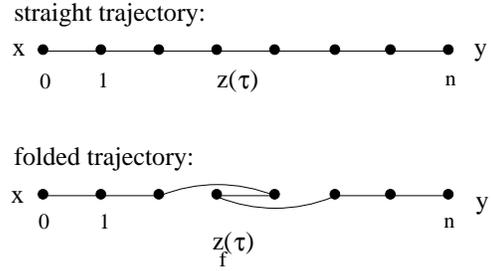}
%\includegraphics[width=2.5in]{folded}
%captionstyle{normal}
\caption{It is possible to create particle-antiparticle pairs using 
folded  trajectories. However folded trajectories are suppressed by the
kinematics, as discussed in the text.}
\label{folded.fig} 
\end{figure}

We now turn to a discussion of the effect of $Z$-diagrams.  For the simple
estimate of the kinetic energy, Eq.~(\ref{KE}), we chose
integration points $z_i=iT/N$ uniformly spaced along a line.  The classical
trajectory connects these points without doubling back, so that they increase
monotonically with proper time, $\tau$.  However, since the integration over each
$z_i$ is independent, there also exist trajectories where $z_i$ does not increase
monotonically with $\tau$.  In fact, for every choice of integration points $z_i$
there exist trajectories with $z_i$ monotonic in $\tau$ and trajectories with
$z_i$ non-monotonic in $\tau$.  The latter double back in time, and describe
$Z$-diagrams in the path integral formalism.   Two such trajectories that pass
through the {\it same\/} points $z_i$ are shown in Fig.~\ref{folded.fig}.  These
two trajectories contain the same points, $z_i$, but ordered in different ways, and
both occur in the path integral. 

Now, since the total self energy is the sum of potential
contributions $V[z,s]$ from all $(z_i,z_j)$  pairs, irrespective of how these
coordinates are ordered, it must be the same for the straight trajectory $z(\tau)$
and the folded trajectory $z_f(\tau)$:
\be
V[z_f,s]=V[z,s]\, .
\ee
However, according to Eq.~(\ref{kin}), the  kinetic energy of the folded
trajectory is larger than the kinetic energy of  the straight trajectory
\be
K[z_f,s]>K[z,s]\, ,
\ee
because it includes some terms with larger values of $(z_i-z_{i-1})^2$.  Since
the kinetic energy term is always positive, the folded trajectory
($Z$-graph) is always suppressed (has a larger exponent) compared with a
corresponding  unfolded trajectory (provided, of course, that $g^2<g^2_{\rm
crit}$).

\begin{figure}
\setcaptionmargin{5mm}
%\onelinecaptionsfalse
\includegraphics[width=2.0in]{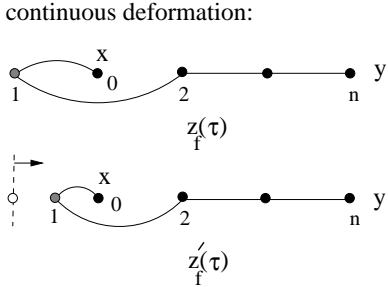}
%\includegraphics[width=2.0in]{folded_deform}
%captionstyle{normal}
\caption{A folded trajectory at the end point of the path, and a similar one
with $z_1$ closer to $z_0$.}
\label{folded2.fig} 
\end{figure}

This argument holds only for cases where the trajectory does {\it not\/} double
back to times {\it before\/} $z_0=0$ or {\it after\/} $z_N=T$.  An example of
such a trajectory is shown in Fig.~\ref{folded2.fig} (upper panel).  Here we 
compare this folded trajectory to another folded trajectory, $z'_f$, with point
$z_1$ {\it closer\/} to the starting point $z_0$ (lower panel of
Fig.~\ref{folded2.fig}).  This new folded trajectory has points spaced closer
together, so that the kinetic energy is smaller and the potential energy is
larger, and therefore  
\be
K[z_f,s]-V[z_f,s] > K[z'_f,s]-V[z'_f,s]\, . 
\ee
It is clear that the larger the folding in the trajectory, the less energetically
favorable is the path, and the most favorable path is again an unfolded trajectory
with no points outside of the limits $z_0<z_i<z_N$.  

While these arguments have been stated in 0+1 dimensions for simplicity, they are
not dependent on the number of dimensions, and can be extended to the realistic
case of 1+3 dimensions.

We conclude that a calculation in quenched
approximation, where the  creation of particle-antiparticle pairs can only come
from $Z$-graphs, must be {\it more\/} stable (produce a larger mass) than a similar
calculation without {\it any\/}
$\chi\bar{\chi}$ pairs.  The quenched
$\chi^2\phi$ theory therefore is bounded by the same limits given in
Eq.~(\ref{gcrit}).  This conclusion supports, and is supported by, the results of
Refs.~\cite{Savkli:1999me,Savkli:2000hw,Savkli:1999gq} which show, in the quenched
approximation, that the $\chi^2\phi$ interaction  is stable for a finite  range of
coupling strengths.

It is now clear that the instability of $\chi^2\phi$ theory must be
due to either (i) the possibility of creating an infinite number of closed
$\chi\bar{\chi}$ {\it loops\/}, or (ii) the presence of an infinite number of
matter particles (as in an infinite medium).  Indeed, the original proof
given by Baym used the possibility of loop creation from the vacuum to prove that
the vacuum was unstable. 

These results provide justification for the stability of relativistic one
boson exchange models that usually exclude matter loops but may
include $Z$-diagrams of all orders. Our argument cannot be easily
extended to symmetric $\phi^3$ theories where it is impossible to make a clear
distinction between $Z$-diagrams and loops. 
 
\section{Conclusions}
In this paper we have given a summary results for scalar interactions obtained with the use of the FSR representation . 
The FSR approach uses a covariant path integral representation 
for the trajectories of particles. Reduction of field theoretical path 
integrals to path integrals involving particle trajectories reduces the 
dimensionality of the problem and the associated computational cost. 

Applications of the FSR approach to 1 and 2-body problems in particular shows 
that uncontrolled approximations in field theory may lead to significant 
deviations from the correct result. Our results indicate that 
use of the Bethe-Salpeter equation in ladder approximation to solve the 2-body bound state problem is a poor approximation.   For the scalar theories examined here, a better approximation to the 2-body problem is obtained using the Gross equation in ladder approximation.   Similarly, use of the rainbow approximation for the 1-body problem gives a poorer result that simply calculating the self energy to second order!  In all of these cases, the explanation for these results seems to be that the crossed diagrams (such as crossed ladders) play an essential role, canceling contributions from higher order ladder or rainbow diagrams.

\begin{acknowledgments}
 We thank Prof. Yu. Simonov for his leadership in this field, and for many useful discussions.  It is a pleasure to contribute to this volume celebrating his 70th birthday.

This work was supported in part by the DOE
grant DE-FG02-93ER-40762, and DOE contract DE-AC05-84ER-40150 under
which the Southeastern Universities Research Association (SURA)
operates the Thomas Jefferson National Accelerator Facility.
\end{acknowledgments}

\bibliographystyle{maik}
\bibliography{rever}

%\bibliography{references}% Produces the bibliography via BibTeX.

\end{document}